\documentclass[prc,unsortedaddress,superscriptaddress,twocolumn,amssymb,showpacs,floatfix,nofootinbib]{revtex4-1}
\usepackage{graphics,epsfig}
\usepackage{amsmath}
\usepackage{dcolumn}   

\begin{document}
\date{\today}
\title{Algorithm  for the asymptotic nuclear coefficients calculations using phase shift data for charged particles scattering}
\author{Yu. V. Orlov}\email{orlov@srd.sinp.msu.ru}
\affiliation{Skobeltsyn Nuclear Physics Institute, Lomonosov Moscow State
University, Russia}
\author{B. F. Irgaziev}\email{irgaziev@yahoo.com}
\affiliation{National University of Uzbekistan, Tashkent,Uzbekistan}
\affiliation{GIK Institute of Engineering Sciences and Technology, Topi, Pakistan}
\author{Jameel-Un Nabi}
\affiliation{GIK Institute of Engineering Sciences and Technology, Topi, Pakistan}

\begin{abstract}
A new  algorithm for the asymptotic nuclear coefficients calculation, which we call the $\Delta$-method, is proved and developed. This method was proposed in Ref. [O. L. Ram\'irez Su\'arez and J.-M. Sparenberg, arXiv: 1602.04082 [nucl-th] (2016)] but no proof was given. We apply it to the bound state  situated near the channel threshold when the Sommerfeld parameter is quite large within the experimental energy region. As a result, the value of the conventional effective-range function $K_l(k^2)$ is actually defined by the Coulomb term.  One of the resulting effects is the wrong description of energy behavior of the elastic scattering phase shift $\delta_l$ reproduced from the fitted total effective-range function $K_l(k^2)$. This leads to an improper value of the asymptotic normalization coefficient (ANC) value. No such problem arises if we fit only the nuclear term.  The difference between the total effective-range function and the Coulomb part at real energies is the same as the nuclear term. Then we can proceed using just this $\Delta$-method to calculate the pole position values and the ANC. We  apply it to the vertices $^4\rm{He}+ {^{12}\rm{C}}\longleftrightarrow {^{16}\rm{O}}$ and $^3\rm{He}+ {^4\rm{He}}\longleftrightarrow {^7\rm{Be}}$. The calculated ANCs can be used to find the radiative capture reaction cross sections of the transfers to the $^{16}\rm{O}$ bound final states as well as to the $^7\rm{Be}$.

\end{abstract}
\maketitle

\section{Introduction}\label{intr}
It is known that many reactions in supernova explosions proceed through subthreshold bound states and low-energy resonance states above the threshold. The single-channel approach is applied to describe these states. To calculate the rate of such reactions, one needs to find the asymptotic normalization coefficient (ANC) of the radial wave function for bound and resonance states, which can be used to calculate radiative capture cross-sections. This process is one of the main sources of the creation of new elements. The reaction $^{12}\rm{C}(\alpha,\gamma){^{16}\rm{O}}$ is considered to be one of the key nuclear processes in stellar nucleosynthesis. The triple alpha and $^{12}\rm{C}(\alpha, \gamma)^{16}\rm{O}$ reaction rates determine the carbon to oxygen ratio at the  completion of helium burning in stars, which, in turn, influences the later stellar burning stages.
An interesting though quite involved paper was published in Ref. \cite{NolettWiringa2011}  where it was shown that  microscopic \textit{ ab initio} calculations of a bound state function are now feasible.

The ANC method was explored as an indirect experimental method for determining the cross sections of peripheral reactions at low energy \cite{muk01}. There are several methods of deriving the bound state ANC from experimental data (see \cite{muk07, akram99}) and references therein).
Recently, the effective-range expansion method has been developed to find the ANC for bound states from an elastic scattering phase-shift analysis (\cite{OrlIrgNik, spren} and references therein).
A renormalized scattering amplitude taking into account the Coulomb interaction was introduced by J. Hamilton \textit{ et al.} \cite{Hamilton} to get the analytic continuation to negative energies.  An important step to calculate the ANC was first taken by Iwinski \text{et al.} \cite{Iwinsky} who discussed a radiative capture process using the Pad\'e-approximant. We note that sufficiently precise measurements of elastic scattering phase-shifts can give crucial information concerning the ANC.

We prove the correctness of the new ANC calculation algorithm based on the equations of the   effective-range theory, as proposed in Ref. \cite{spren16} (see also \cite{blokh16}) which we call the $\Delta$-method. This $\Delta$-method allows us to avoid some problems arising when charges of colliding particles increase.  We note that the authors of \cite{spren16} call the $\Delta$-method ``an approximation''. The authors of \cite{blokh16} describe the effective-range function in a model with ``rectangular nuclear potential plus the Coulomb interaction'', and conclude that their results ``justify'' the $\Delta$-method.   In the present paper, we consider some limitations of the conventional procedure for doing this in the frame of the effective-range expansion (ERE), as well as of the Pad\'e approximation.

First of all, we point out that we use a one-channel and one-particle approximation, which ignores the inner structure of colliding nuclei. Consequently, the ERE cannot be applied, for example, to nuclear levels of a collective type. Moreover, this approach has an inner contradiction: the ERE results for a bound state can be trusted more when the experimental phase-shifts are known at energies which are as small as possible. But, due to the Coulomb repulsion barrier, the experimental phase-shift uncertainty increases while the energy decreases. As a result, for some nuclear systems one cannot  deduce ANC values which are precise enough when applying this theory by using existing experimental phase-shift data. A necessary check should give a reasonable reproduction of the energy pole position for a given bound state. Usually this energy is known with better precision than for phase-shifts, so the pole position can be added to the input data. Note that the position of the pole corresponds to the general condition
\begin{equation}\label{zero}
\cot\delta_l-i=0.
\end{equation}
which is the same for charge-less particle collision. This well-known  pole condition can be seen from the expression for the renormalized amplitude of the elastic scattering
\begin{equation}\label{fl2}
\tilde{f}_l(k)=\frac{1}{k(\cot\delta_l-i)\rho_l(k)},
\end{equation}
where the function $\rho_l$ is defined by the equation
\begin{equation}\label{rho}
\rho_l(k)=\frac{2\pi\eta}{e^{2\pi\eta}-1}\prod_{n=1}^l\Bigl(1+\frac{\eta^2}{n^2}\Bigr).
\end{equation}
Here $\delta_l$ is the nuclear phase shift modified by the Coulomb interaction, and
$\eta=\xi/k$ is the Sommerfeld parameter, $\xi=Z_1Z_2\mu\alpha$,
$k=\sqrt{2\mu E}$ is the relative momentum, $\mu$, $E$  are the reduced mass, the relative energy  of the colliding nuclei with the charge numbers $Z_1$ and $Z_2$,  respectively, and $\alpha$ is the fine-structure constant.
Writing the expression $\cot\delta_l$ in a non-physical energy region in this equation and elsewhere, we mean its analytical continuation, since the phase shift is defined only in the positive energy region.

Here, we formulate and develop a new algorithm for the ANC calculation which is mainly independent of the Coulomb interaction. The point is that in the physical energy region the effective-range function is real and presents the sum of the  nuclear and the Coulomb  terms. The expression for $\cot\delta_l(E) $ at $l=0$ which is valid for energy $E$ in the physical domain is given in  Landau and Lifshitz's textbook (see Eq. (136.11) in \cite{Landau}). Only $\cot\delta_l$  provides information about the nuclear interaction. The other components depend only on the Coulomb Sommerfeld parameter and are therefore known exactly.  This fact leads to a new calculation algorithm: one needs to fit only the first nuclear term  $C_0^2\cot\delta$ in (\ref{CoulombKl}) below, while the conventional algorithm consists of fitting the whole effective-range function $K_l(k^2)$.

The example of the $\alpha {^{12}\rm{C}}$ bound system with $J^\pi =1^-$ level near the threshold shows that the Sommerfeld parameter is too big in the experimental region of the measured phase shift.  In fact, the whole $K_l(k^2)$ function is actually numerically defined by the second Coulomb term. Consequently, the nuclear term only slightly corrects $K_l(k^2)$. This results in a wrong restoration of the phase-shift energy dependence when one tries to reproduce the experimental phase shift from the fitted effective-range function $K_l(k^2)$ and an improper ANC value.

No such problem arises when fitting only the nuclear term, which is the difference ($\Delta_l$)  defined below in (\ref{Delta-R}). We show that $\Delta_l$ must be zero at the negative energy of a bound pole.  This  enables us to deduce the residue and other calculating constants in an explicit form
(see below).

The pole condition $\cot\delta_l = i$ must be fulfilled for a bound state. If it is not realized approximately with parameters found without including binding energy into the initial  data, then one can conclude that either the considered state is not one-particle, or the phase-shift data is not accurate enough in the low-energy region. In the latter case, it is reasonable to include the experimental pole position, i.e. the binding energy, into the input data in order to clarify the fitting parameters.

Besides, a fitting series should be convergent at the energy region near the pole. As was pointed out in \cite{Orlov et al.}, the convergence radius in the complex energy plane is determined by the Feynman diagram of the scattering process, which has a singularity situated nearest to the zero energy when this zero energy is chosen as the centered point of an the expansion in the powers of the energy. This is due to the fact that such a singularity has a logarithmic behavior which is not present in the analytic form of the effective-range or Pad\'e approximation scattering amplitude.

We believe that finding the nuclear vertex constant (NVC) and ANC of bound states by using an experimental phase-shift analysis (see \cite{OrlIrgNik, spren} and references therein) is better than  other methods within the model described above.

The article is organized as follows. In Sec. II we present the main formulas for proving the $\Delta$-method. The final equations presented can be used to deduce the nuclear vertex constant (NVC) or $G_l$, the residue $W_l$ and the ANC.

In Sec. III we apply the $\Delta$-method to the bound states of $^{16}\rm{O}$ situated below the $\alpha+^{12}\rm{C}$ threshold. These include the bound $^{16}\rm{O}$ states with $J^\pi =0^+$ ($\epsilon_1=7.162$ MeV for the ground state and $\epsilon_2=1.113$ MeV for the excited state), $J^\pi =1^-$ ($\epsilon=$0.045 MeV) and $2^+$ ($\epsilon=$0.245 MeV). We use as input data the phase-shifts from the R-matrix analysis augmented by the binding energies of the states considered. We compare some of our results with those which were published in the literature, in particular, in \cite{spren16}.

In Sec. IV we apply the $\Delta$-method to the ground and the first excited states of the $^7\rm{Be}$ nucleus when $J^\pi=3/2^-$ and $J^\pi=1/2^-$. We use the experimental phase-shifts for the $^3\rm{He}^4\rm{He}$ scattering and the binding energies known in the literature. Due to larger experimental phase-shifts uncertainties compared with $\alpha^{12}\rm{C}$ data we find only some intervals for the ANC and other constants of the $^7\rm{Be}$ bound states.

In Sec. V (conclusion) the results of the present paper are discussed.
We show that the $\Delta$-method is strictly valid for calculating the ANC in the frame of the ERE theory. In fact, this constitutes a final step in the formulating the ERE approach, which simplifies calculations and excludes the unnecessary details of the Coulomb interaction. The $\Delta$-method is necessary for large charges of colliding nuclei when the conventional ERE approach becomes invalid. The results of the paper are important for nuclear astrophysics and for the theory using Feynman diagrams to describe the amplitude of the direct nuclear reaction.

In the following we use the unit system $\hbar=c=1$.

\section{The  $\Delta$ algorithm of the effective-range function expansion method for bound  states}\label{eff-range}

In this section, some relationships in the effective-range expansion method (ERE) are given.

The renormalized scattering amplitude is written as
\begin{equation}\label{renormalized amplitude}
 \tilde f_l (k) = \frac{k^{2l}}{K_l(k^2)-2\xi D_l(k^2)h(\eta)}
\end{equation}
(see, for example, \cite{OrlIrgNik} and definitions below), where the effective-range function $K_l(k^2)$
 borrowed from Ref. \cite{haer} has the form:

\begin{equation} \label{CoulombKl}
K_l(k^2) = 2 \xi D_l(k^2)\left[C_0^2(\eta)(\cot\delta_l - i)+h(\eta )\right],
\end{equation}
where
\begin{eqnarray}
C_0^2(\eta)&=&\frac{\pi} {\exp(2\pi\eta)-1},\label{C02}\\
h(\eta)&=&\psi(i\eta) + \frac{1}{2i\eta} - \ln(i\eta),\label{h-eta}\\
D_l(k^2)&=&\prod_{n=1}^l\Bigl(k^2+\frac{\xi^2}{n^2}\Bigr),\qquad D_0(k^2)=1,\label{DL_k}
\end{eqnarray}
and $\psi(x)$ is the digamma function. We note that the
effective-range function  $K_l(k^2)$ is real and continuous on the positive real axis of the
energy plane and can be analytically continued from the real positive axis to pole energies situated in the complex plane of the momentum $k$ or energy $E$.

 We define the $\Delta_l(k^2)$ function as

\begin{equation}\label{Delta-R}
\Delta_l(k^2)=\frac{\pi}{\exp(2\pi\eta)-1}\cot\delta_l,
\end{equation}
on the positive energy semi-axis.
We see that $\Delta_l(k^2)$  can be considered as an analytical function of  $k^2$ with the possible exception of a few poles.
The factor in front of $\cot\delta_l$ depends on the Sommerfeld parameter $\eta$.  This factor is quite important because it deletes the essential singularity of $\cot\delta_l$ at the zero energy. This  singularity follows from the  $\delta_l$ energy behavior   near the origin due to the Coulomb interaction (see Eq. (6) in \cite{orlov16} derived in  \cite{Mur83}).

The effective-range function $K_l(k^2)$ used in the conventional method  can be analytically continued to
the nonphysical Riemann energy surface, and particularly to the negative energy region. On the negative energy semi-axis, the Coulomb function $h(\eta)$ defined by Eq. (\ref{h-eta}) is written as
\begin{equation} \label{H-eta-neg}
h(-i\mid\eta\mid)= \psi(\mid\eta\mid) +\frac{1} {2\mid\eta\mid} - \ln\mid\eta\mid,
\end{equation}
because $k\to i\kappa$ and $\eta\to -i\mid\eta\mid=-i\xi/\kappa$ when we change $E\to -\mid E\mid$ ($\kappa=\sqrt{2\mu\mid E\mid})$.
This equation means that $ h(\eta)$ is a real function of $E$ when $E<0$.
For $E>0$ one can write using  Eq. (9) from Ref. \cite{OrlIrgNik}  (see formulas (6.3.17) for the digamma function in \cite{abramowitz}) the following equation

\begin{equation} \label{H-eta-arviex}
h(\eta)= \frac{i\pi}{\exp(2\pi\eta)-1}+\eta^2 \sum_{n=1}^\infty\frac{1}{[n(n^2+\eta^2)]}-\ln\eta-\zeta,
\end{equation}
where $\zeta \simeq $ 0.5772 is the Euler constant.
Comparing Eqs. (\ref{CoulombKl}), (\ref{h-eta}) and (\ref{H-eta-arviex})  we see that the whole sum $K_l(k^2)$ is real for positive energies ($E>0$) because  the imaginary part (the first term in (\ref{H-eta-arviex})) exactly cancels out that in (\ref{CoulombKl}) and the rest in (\ref{H-eta-arviex}) can be denoted as $\Re h(\eta )$. So, in the physical energy region, in Eq. (\ref{CoulombKl}) the factor in square brackets  of the effective-range function $K_l(k^2)$ may be written as the sum of  the  nuclear and  Coulomb parts:

\begin{equation} \label{CoulombKl-pos}
K_l(k^2) = 2 \xi D_l(k^2)\left[\Delta_l(k^2)+ \Re h(\eta)\right].
\end{equation}

\begin{figure*}[thb]
\begin{center}
\parbox{12.0cm}{\includegraphics[width=12.0cm]{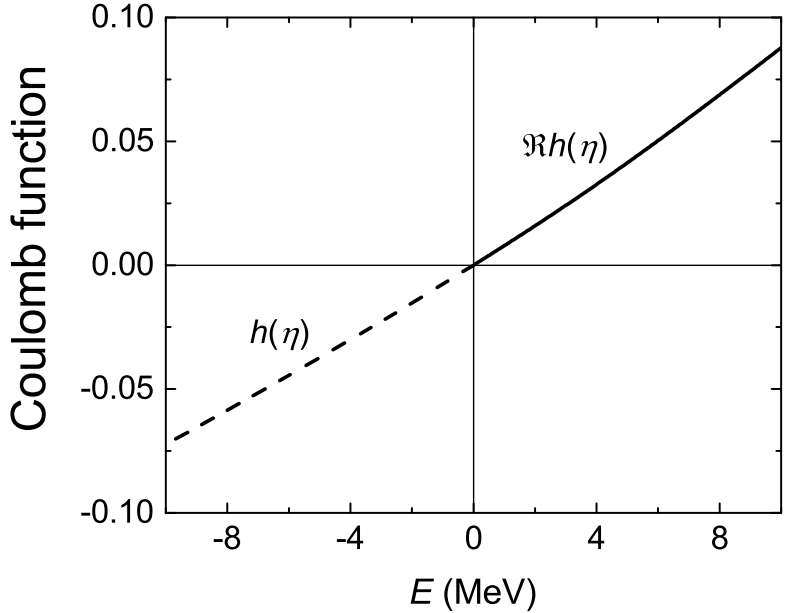}}
\caption{Coulomb part $\Re h(\eta)$  in the physical energy region (real, positive $E$) and $h(\eta)$ in the bound-state energy area (real, negative $E$). \label{fig1}}
\end{center}
\end{figure*}	
In Fig. \ref{fig1} we show how the real  function $\Re h(\eta )$ given in the region $E>0$ continues to the real function $h(\eta)$ in the region $E<0$.
Actually, the curve in Fig. \ref{fig1} is one function of  energy $E$ along the real axis which looks like a straight line around zero presented by the function (see \cite{Hamilton} and Eqs. (6.3.18, 6.3.19) in \cite{abramowitz})   $1/12 \eta^2 \sim E$ which has  equal limits for $E \to \pm 0$ and  the same  for the derivative of this function.

So we can recast Eq. (\ref{renormalized amplitude}) for the renormalized amplitude continuation from positive to negative energies as
\begin{equation}\label{renormalized amplitude with Delta}
 \tilde f_l (k) = \frac{k^{2l}}{2\xi D_l(k^2)\Delta_l(k^2)}.
\end{equation}

The analytical continuation to a negative energy of the first term (the nuclear part) in Eq. (\ref{CoulombKl-pos}) is realized by fitting $\Delta_l(k^2)$ in the experimental domain. To do this, we use polynomial or Pad\'e approximations, which are also the real functions of the real $E$, both at $E>0$ and $E<0$.

Our main aim is to put on a rigorous basis a new algorithm for the ANC calculation for bound states, which we call the $\Delta$-method.  We analyze the  multiplier $F_l=C_0^2 (\eta)(\cot\delta_l - i)+h(\eta)$ of the effective-range function $K_l (k^2)$, which is the sum of the nucleus and Coulomb ($h(\eta)$) parts. In the physical region, functions $F_l$ and $C_0^2 \cot\delta_l$ are real because, as we show above,
$\Im(h(\eta$))  compensates for $-iC_0^2(\eta)$. Our definition of the nuclear part $\Delta_l=C_0^2 (\eta)(\cot\delta_l)$ differs slightly from those in \cite {spren16} and \cite {blokh16}  because we do not include all the other multipliers. This does not change the main results.
Since the functions $F_l$ and $\Delta_l$ are real in the physical energy region, their continuation to the negative energy are also real functions. But in the area $E<0$, the effective-range function includes
$F_l = C_0^2(\eta)(\cot\delta_l-i)+h(\eta)$, where $h(\eta)$  is the real Coulomb function. A bound-state pole at the negative energy is defined in the conventional method by the condition $F_l-h(\eta)=C_0^2(\eta)(\cot\delta_l-i)=0$. So the only possibility of obtaining $\cot\delta_l-i = 0$ is to find an energy when $\Delta_l = 0$. In this case, there is a pole at a bound-state energy. This is because, according to quantum mechanics, the continuous energy spectrum exists in the physical region $E > 0$, and a discrete spectrum arises when $E < 0$.  In fact, the $\Delta$-method is the last step in the effective-range approach, which simplifies a calculation and its analysis. Besides, it enables the application of the $\Delta$-method to systems where colliding particles have larger charges.

The consideration given above  proves the correctness of the $\Delta$-method.
Using this method, we can start finding concrete constants.
 Let us assume that the experimental phase shift does not pass through zero or $\pi n$ at the energy interval where the scattering phase shift is measured.  Then we can expand $\Delta_l(k^2)$ by a polynomial:
\begin{equation}\label{polynomial}
\Delta_l(k^2)=a_0+a_1 k^2+a_2 k^4+\cdots,
\end{equation}
where the coefficients of the expansion are determined by fitting the known scattering phase shift from the experimental data. The number of terms in the polynomial (\ref{polynomial}) depends on obtaining the best fit result. This number is not limited in the convergence energy region.

The pole position corresponding to the bound state is known with high precision. Therefore it is reasonable to include it in the array of the experimental data. Note that the function chosen as a polynomial or Pad\'e  approximation with fitted parameters must pass through this pole, or at least be close to it, after inserting the fitted parameters into the expressions for calculating the ANC of the bound state wave function with high accuracy. Therefore, it is reasonable to single out the factor $1+k^2/ \kappa_b^2$ ($\kappa_b^2=2\mu\epsilon_b$, $\epsilon_b$ is a binding energy) in Eq. (\ref{polynomial}), taking into account that the pole of the amplitude is simple and single. The inclusion of a specific binding energy in the function $\Delta_l(k^2)$ severely limits the number of possible options for fitting, and the inclusion of two binding energies should obviously restrict the fitting further.
So, after including one bound state, we rewrite Eq. (\ref{polynomial}) as
\begin {equation} \label{DeltaPole-1}
\Delta_l(k^2)=(1+k^2/\kappa_b^2) \Phi_l(k^2),
\end{equation}
where the function $\Phi_l(k^2 )$ may be fitted by a polynomial or a Pad\'e approximation using an experimental phase-shift data.

To deduce the residue $W_l$,  we can replace $\Delta_l(k^2)$ in the denominator of the renormalized  scattering amplitude $\tilde{f}_l(k)$  of charged particles Eq. (\ref{renormalized amplitude with Delta}) by the polynomial (\ref{DeltaPole-1})
(or Pad\'e-approximant) extended to the negative energy region up to the point where $\Delta_l(k^2)=0$,  and obtain
\begin{eqnarray}
\tilde{f}_l(k)&=&\frac{k^{2l}\kappa_b^2}{2\xi D_l(k^2)(k^2+\kappa_b^2) \Phi_l(k^2)}=\nonumber\\
&& \frac{k^{2l}\kappa_b^2}{2\xi D_l(k^2)(k-i\kappa_b)(k+i\kappa_b)\Phi_l(k^2)}.\label{tilde-f}
\end{eqnarray}
From Eq. (\ref{tilde-f}), the residue takes the form
\begin{eqnarray}
W_l &=&\lim_{k^2 \to -\kappa_b^2}\frac{k^{2l}\kappa_b^2}{2\xi D_l(k^2)(k+i\kappa_b)\Phi_l(k^2)}=\nonumber\\
&&\frac{i(-1)^{l+1}\kappa_b^{2l+1}}{4\xi D_l(-\kappa_b^2)\Phi_l(-\kappa_b^ 2)}.\label{res1}
\end{eqnarray}
Note that the residue $W_l$ is dimensionless.
If the amplitude has a greater number of bound states, then  Eq. (\ref{res1}) should be applied to  every bound pole.

According to the known relations between the NVC ($G_l$),  the residue $W_l$ and ANC we can use the following equations
(see, for example, Ref. \cite{ECHAYA}):

\begin{equation}\label{NVC}
G_l^2=-\frac{2\pi \kappa_b}{\mu^2}W_l,
\end{equation}
and
\begin{equation}\label{Abs(ANC)}
C_l=\frac{\mu}{\sqrt{\pi}}\frac{\Gamma(l+1+\eta_b)}{l!}\mid G_l\mid,
\end{equation}
where $\eta_b=\xi/\kappa_b$.

\section{Application of the $\Delta$-method to the bound states of $^{16}\rm{O}$ situated  below the $\alpha+{^{12}\rm{C}}$ channel threshold}

First, we analyze the levels of  $^{16}\rm{O}$ nucleus  below the $\alpha+^{12}\rm{C}$ threshold presented in \cite{nucldata}. We see that the levels with  energies and  quantum numbers $E_x=0,\, 6.0494\pm 1.0\, (J^\pi=0^+),\,\, 6.9171\pm 0.6,\,(J^\pi=2^+)$ and $7.11685\pm 0.14\, (J^\pi=1^-)$ can be treated as bound states of $\alpha$-particle and $^{12}\rm{C}$.

Therefore, we apply the algorithm derived above to the bound $^{16}\rm{O}$ states with $J^\pi =0^+$ ($\epsilon_{1})=7.162$ MeV for the ground state and $\epsilon_{2}=1.113$ MeV for the excited state), $J^\pi =1^-$ ($\epsilon=$0.045 MeV) and $2^+$ ($\epsilon=$0.245 MeV). Here and below we omit index $b$ in the designation of a binding energy $\epsilon_b$.  As clearly seen from Figs. 9-11 of the paper \cite{tisch}, the phase-shifts of the $\alpha {^{12}\rm{C}}$ scattering in the states $J^\pi =0^+,\,1^-$ are well-defined in the energy range $0<E_{lab}<6,\, \rm{MeV}$ in the lab system and do not contain resonance states. Therefore, we can use a polynomial expansion for the fitting function $\Delta_l(k^2)$. However, in the state $J^\pi=2^+$, two very narrow resonances exist as well as two zeros
in the corresponding energy phase-shift behavior.  Therefore, we take the two-pole Pad\'e-approximant as a fitting $\Delta_l(k^2)$ function.  So, to calculate the NVC and ANC corresponding to the bound  states of $^{16}\rm{O}$ below $^{16}\rm{O}\to \alpha+{^{12}\rm{C}}$ threshold, we take the fitting functions $\Delta_l(k^2)$, which depend on the relative $\alpha {^{12}\rm{C}}$ energy as
\begin{eqnarray}
\Delta_0(k^2(E))&=&\Bigl(1+\frac{E}{\epsilon_1}\Bigr)\Bigl(1+\frac{E}{\epsilon_2}\Bigr)(a_0+a_1 E+a_2 E^2),\label{Delta_0}\\
\Delta_1(k^2(E))&=&\Bigl(1+\frac{E}{\epsilon_1}\Bigr)(a_0+a_1 E+a_2 E^2),\label{Delta_1}\\
\Delta_2(k^2(E))&=&\frac{1+E/\epsilon_1}{\Bigl(1-\frac{E}{E_{z1}}\Bigr)\Bigl(1-\frac{E}{E_{z2}}\Bigr)}
\times\nonumber\\
&&(a_0+a_1 E+a_2 E^2+a_3 E^3),\label{Delta_2}
\end{eqnarray}
in $0^+,\,\,1^-,\,\, 2^+$ states respectively. Here $\epsilon_i$ is the corresponding binding energy in the considered states, $E_{z1}=2.636 \,\rm{MeV}$, $E_{z2}=3.980\,\rm{MeV}$ are the energies of the zeros of the partial scattering amplitude in the $2^+$ state. For a brevity we omit in (\ref{Delta_0}), (\ref{Delta_1}) and (\ref{Delta_2}) the index $l$. In fact, the sets of the fitted coefficients $a_i$ in the Eqs. above are really different for the different $l$.

We find the parameters of the  fitting $\Delta_l$ functions by applying the phase-shift data presented in Ref. \cite{tisch}, which was obtained through an R-matrix fit of high-precision cross sections measured at the energy interval of the region which are close to the $\alpha+{^{12}\rm{C}}$ threshold. The fitting forms of the $\Delta_l(k^2(E))$ functions given above in Eqs. (\ref {Delta_0}) - (\ref{Delta_2}) are quite sufficient to describe the experimental phase-shifts for the selected states and can be used for calculating the ANC and other constants.
We note that here and elsewhere we also omit index of a state in $k_b^2$.

\begin{figure*}[thb]
\begin{center}
\parbox{12.0cm}{\includegraphics[width=12.0cm]{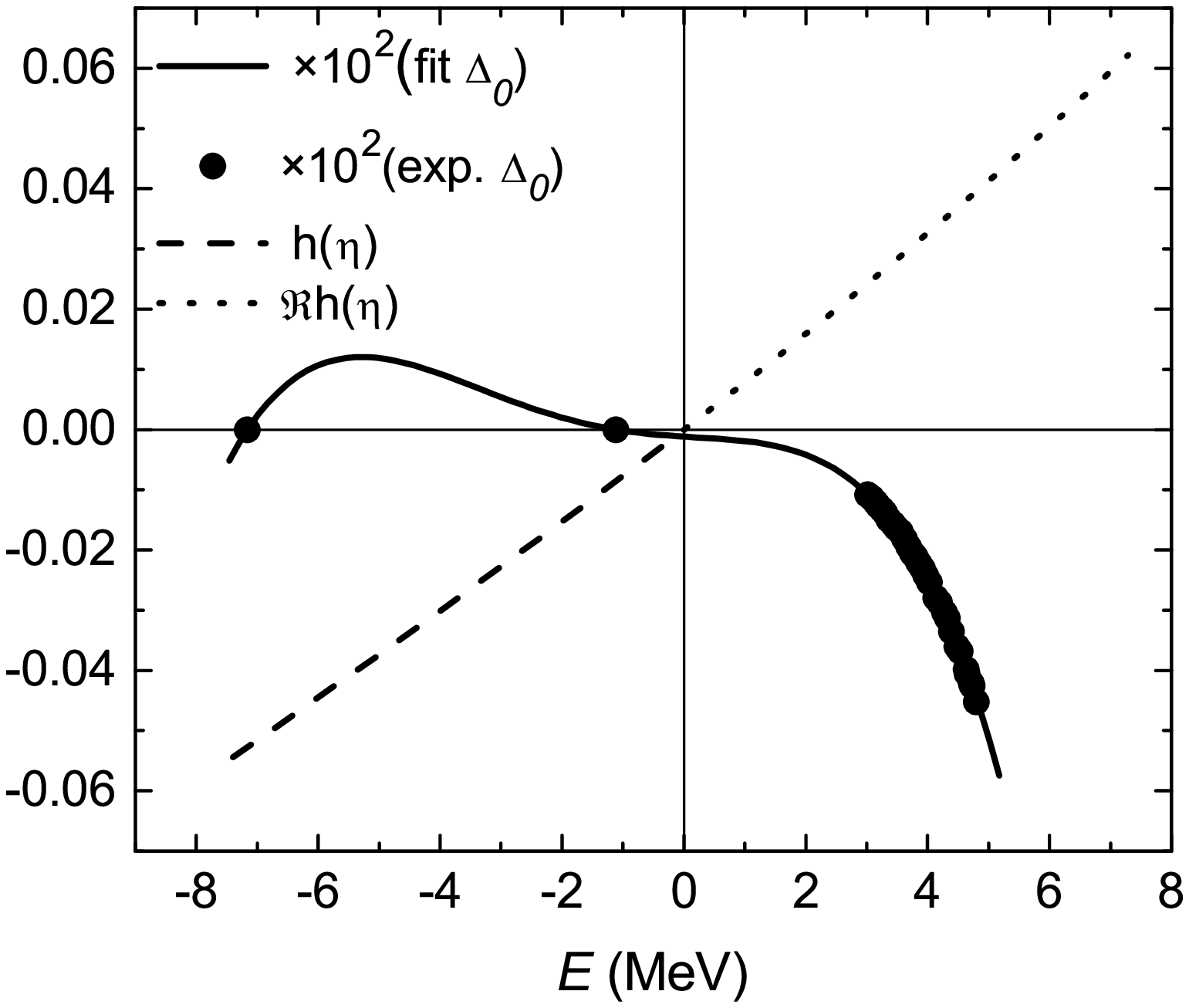}}
\parbox{12.0cm}{\includegraphics[width=12.0cm]{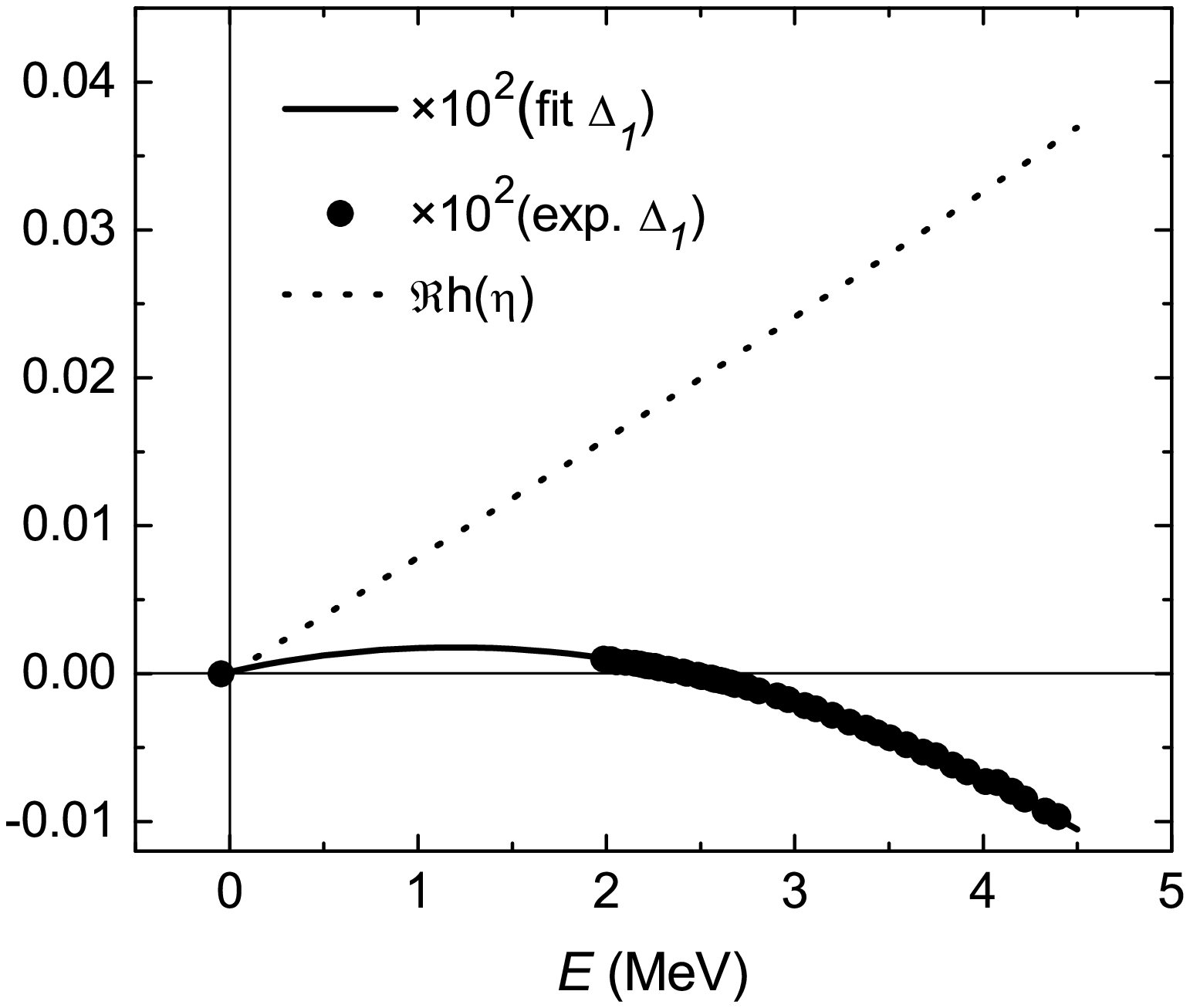}}
\caption{Dependence of the fitted $\Delta_l(k^2(E))$ functions determined by Eqs. (\ref{Delta_0}) (top), (\ref{Delta_1}) (bottom),  $\Delta_l(k^2(E))=\frac{\pi}{\exp(2\pi\eta)-1}\cot\delta_l$  and the real part of  the Coulomb function $h(\eta)=\psi(i\eta)+\frac{1}{i2\eta}-\ln(i\eta)$ on the relative energy of $\alpha {^{12}\rm{C}}$ elastic collision in $J^\pi=0^+\,\,\rm{and}\,\,1^-$ states. The experimental data are taken from Ref. \cite{tisch}.  \label{fig2}}
\end{center}
\end{figure*}
\begin{figure*}[thb]
\begin{center}
\parbox{12.0cm}{\includegraphics[width=12.0cm]{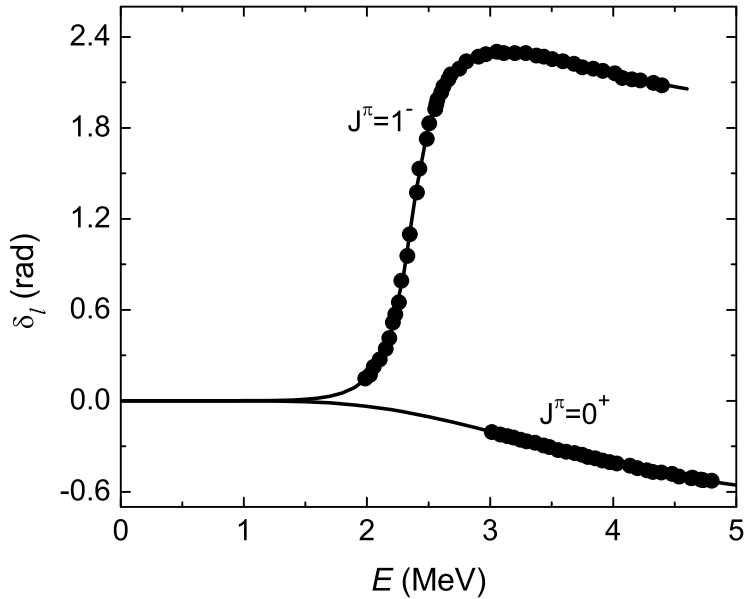}}
\caption{Comparison of the restored phase-shift (solid line) with the  experimental phase-shift data (filled circles) of $\alpha {^{12}\rm{C}}$ elastic collision  in  $J^\pi =0^+,\,1^-$ states. The experimental data are taken from Ref. \cite{tisch}. \label{fig3}}
\end{center}
\end{figure*}

\begin{figure*}[thb]
\begin{center}
\parbox{12.0cm}{\includegraphics[width=12.0cm]{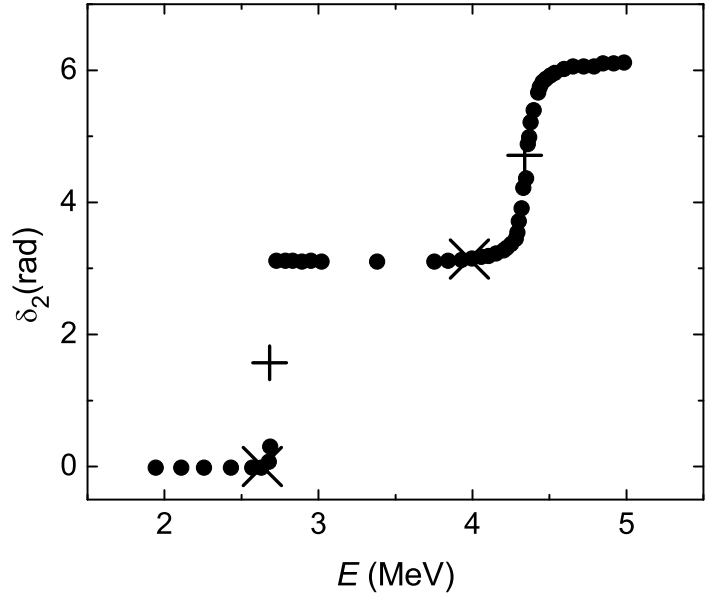}}
\caption{Continuous branch of the $\alpha{^{12}\rm{C}}$ phase shift when $J^\pi=2^+$: experimental phase shift (points), zeros (2 crosses $\times$) and resonances (2 crosses +).The experimental data are taken from Ref. \cite{tisch}.
\label{fig4}}
\end{center}
\end{figure*}

\begin{figure*}[thb]
\begin{center}
\parbox{12.0cm}{\includegraphics[width=12.0cm]{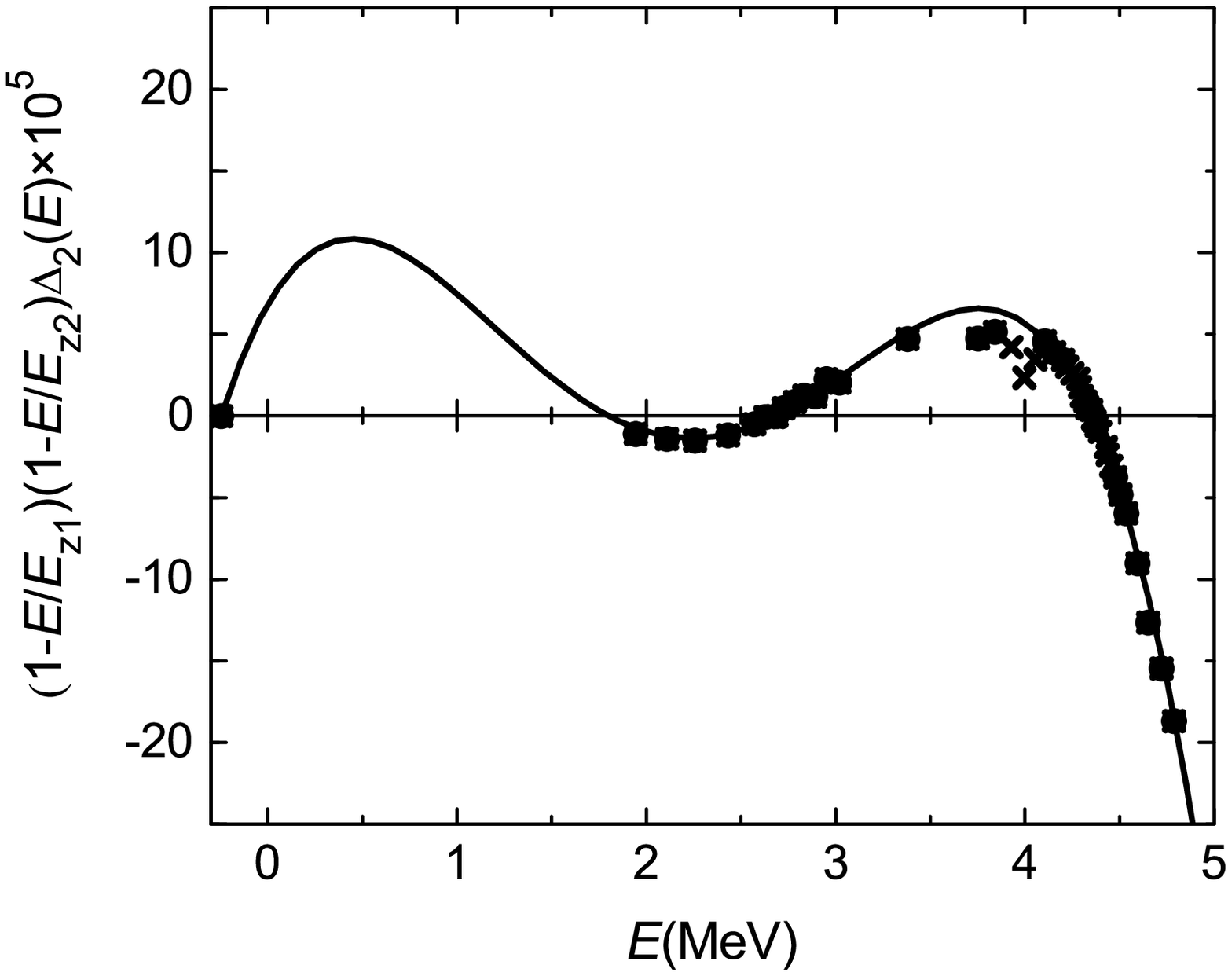}}
\caption{Comparison of  $(1-E/E_{z1})(1-E/E_{z2})\Delta_2(k^2(E))$ function, where $\Delta_2(k^2(E))$ is defined by Eq. (\ref{Delta_2}) with the corresponding experimental data in the $J^\pi=2^+$ state.  The experimental data are taken from Ref. \cite{tisch}. \label{fig5}}
\end{center}
\end{figure*}

A comparison of $\Delta_l(k^2)$ with  $\Re{h(\eta)}$ defined by Eq.(\ref{h-eta}) shows that  $\mid\Delta_l(k^2)\mid$ is much smaller ($\sim 10^2-10^3$ times) than $\Re{h(\eta)}$ in the experimental energy interval (Fig. \ref{fig2}). Therefore, the fitted parameters of the effective-range function (\ref{CoulombKl-pos}) are in this case determined mostly by the Coulomb term $\Re{h(\eta)}$,  and its application leads to an ANC value which is not correct. For example, when we  use the  ERE approach, the phase-shift behavior is restored  incorrectly for the  $\alpha {^{12}\rm{C}}$ collision in $1^-$ state \footnote{J-M Sparenberg points this out to us in a private email. We thank him very much for this observation.}. There is no such problem if we apply the $\Delta$-method. Fig. \ref{fig3} demonstrates a good match of the restored phase shift with the experimental data for the $J^\pi=0^+,\,1^-$ states. We include $\Delta_l(-k_b^2)=0$ into the input data because the binding energies of $^{16}\rm{O}=\alpha+{ ^{12}\rm{C}}$ are known with higher precision.

The specific and rather complex energy behavior of the $\alpha{^{12}\rm{C}}$ scattering phase shift
$\delta_2(E)$ for the $J^\pi=2^+$ state needs the construction of a continuous branch of this experimental phase shift. To obtain this from the experimental data of Ref. \cite{tisch}, we add $\pi$ to the experimental values of  the phase shift when $E>E_{z1}=2.636$ MeV. This leads to the increasing steps of the $\Delta_2(k^2(E))$ (see Fig. \ref{fig4}), which we use to obtain the fitting function $\Delta_2(k^2(E))$. As mentioned above there are two phase-shift crossings of the zero and $\pi$ lines, which means that there are also $\cot \delta_2$ poles at these crossing energies.  It is easy to find their positions ($E_{z1}=2.636$ MeV and $E_{z2}=3.980$ MeV) by interpolating the neighboring experimental  $\Delta_2(k^2(E))$ points. These crossing points are denoted by two crosses ($\times$) in Fig. \ref{fig4}. This leads to the Pad\'e-approximant for $\Delta_2 (k^2(E))$, which is determined by Eq. (\ref{Delta_2}). We achieve an accurate description of the experimental $\Delta_2(k^2(E))$ energy behavior when multiplying $\Delta_2(k^2(E))$ by two factors $(1-E/E_{z1})(1-E/E_{z2})$ to get the polynomial expansion up to $E^3$.
The fitting result for the state $J^\pi=2^+$ is shown in Fig. \ref{fig5}. Table \ref{tab1} shows the calculated constant values related to the $^{16}\rm{O}$ nucleus, which are obtained using the $\Delta$-method.
\begin{table*}[!ht]
\caption{Result for values of the residue ($\mid W\mid$),  NVC ($\mid{G}^2_l\mid$) and
ANC ($\mid C_l\mid$) obtained by fitting the elastic $\alpha {^{12}\rm{C}}$ scattering phase-shifts presented in Ref. \cite{tisch}.}
\begin{ruledtabular}
\begin{tabular}{ccccc}
$J^\pi$&Binding energy $\epsilon$\,(MeV)&$\mid W_l\mid$
&$\mid{G}^2_l\mid$\,(fm)\quad\,\,&ANC ($C_l$) (\rm{fm}$^{-1/2})$\\
\hline
&&&&\\
$0^+$& 7.162& 187.4 &5.950& 21.76\\
$0^+$& 1.113& $4.396\times 10^3$ &55.03& 405.7\\
$1^-$& 0.045& 53.57 &0.1348& $2.073\times 10^{14}$\\
$2^+$& 0.245& 1.152 &0.0068& $5.050\times 10^{4}$  \\
\end{tabular}
\end{ruledtabular}
\label{tab1}
\end{table*}
Here,  ANC=$21.76\,\rm{fm}^{-1/2}$  for $J^\pi=0^+$ (ground state) which is very close to our result obtained by applying the conventional ERE method \cite{orlov16} and  the value ANC=$2.073\times 10^{14}\,\rm{fm}^{-1/2}$ $(J^\pi=1^-)$ is in perfect agreement with the results presented in Refs. \cite{brune,spren16}, but double our result in Ref. \cite{orlov16}. The authors of Ref. \cite{brune} applied the R-matrix method for the fitting the experimental data, whereas in \cite{spren16} the Pad\'e approach was used for the fitting function.  Note that the calculated value of the ANCs  is dependent on how many experimental points are chosen in the energy interval for fitting. Therefore, the deviation of the pole position by several per cent points leads to changes in the order of the ANC  due to the $\Gamma$-factor in Eq. (\ref{Abs(ANC)}),  but the restored phase-shift curve always  agrees with the experimental data. However, the $W_l$ and  $G_l$ do not deviate as much. The behavior of $\Delta_l$ shown in  Fig. 1 (\textit{b, c}) of Ref \cite{spren16} is very different from our curve given in Fig. \ref{fig2} (bottom). In addition, the maximum in $\Delta_1$  of \cite{spren16} is reached at $E=3.7 \,\rm{MeV}$, while our curve with a parabola form has its maximum at $E=1.2\,\rm{MeV}$.  We also check the convergence of the polynomial expansion (\ref{polynomial}) by adding the next high order terms in Eqs. (\ref{Delta_0}, \ref{Delta_1}). The fitted parameter for the fourth term  is  three orders  less than the parameter in the third term. The convergence of the expansions in Ref. \cite{spren16} is poor (see Fig.1. (\textit{b}, \textit{c})) in spite of the fact that the restored phase shift (Fig.1. (\textit{a})) behaves well in the experimental energy range. We also obtain the narrow resonance positions $E_{2r}$ for the state $J^\pi=2^+$ at the following energies: $E_{2r}$=2.685 MeV and 4.349 MeV.  These are in  reasonable agreement with the results obtained in (Table III) of Ref. \cite{irgaz-14} : 2.683 MeV and 4.339 MeV respectively. These points are shown by the symbol (+) in Fig. \ref{fig4}.

We note that the ANC for the 2$^+$ state is quite sensitive to the choice of phase-shift experimental data fitting.
This choice has sometimes been applied while using the conventional method if individual points are strongly displaced compared to neighboring points to get a smoother energy dependence. The curve in the Fig. 5 is obtained without the two points indicated
by the crosses. When  we take the experimental data indicated by crosses in Fig.\ref{fig5},  we obtain  a new set of the values:
\begin{eqnarray}
\mid W_l\mid &=&0.7564, \,  \mid G^2_l\mid=0.0045\,\rm{fm},\, \rm{and}\nonumber\\
\rm{ANC}(C_2)&=&4.082 \times 10^4\,\rm{fm}^{-1/2}.\label{ANC-2}
\end{eqnarray}

\section{The application to the ground and the first excited states of the $^7\rm{Be}$ nucleus }\label{berilium}
The ground and the first excited states of the $^7\rm{Be}$ nucleus in a two-body description  may be treated as the bound states of the nuclei $^3\rm{He}$ and $^4\rm{He}$. The binding energies are $\epsilon_{3/2}=1.587\,\rm{MeV},\epsilon_{1/2}=1.158\,\rm{MeV}$ with the spin-orbital splitting taken into account. Such an assumption is justified by the closeness of the phase-shift behavior of the $^3\rm{He}{^4\rm{He}}$ elastic scattering phase-shifts at a small energy \cite{boykin, hardy} when $J^\pi=3/2^-$ and $J^\pi=1/2^-$ states. In this case, the Coulomb function $\Re{h(\eta)}$ is comparable to the term
$\mid\Delta_l\mid$ of Eq. (\ref{CoulombKl-pos}), but  the terms have opposite signs. Calculations of the ANC for the  $3/2^-$ and $1/2^-$ bound state wave functions were carried out by  several authors \cite{orlov1, spren1, arkadiy, baye1, rakhim} using the different methods. We would like to stress that the solution of the Schr\"{o}dinger equation using the phase equivalent potentials leads to a value of the ANC within quite a large range. Therefore, searching for the ANC by the $\Delta$-method applying the analytical continuation  to the unphysical range is preferable. For the fitting procedure, the experimental phase-shift data from \cite{boykin, hardy} can be used.
Again, we take the fitting function for $\Delta_1(k^2(E))$  as

\begin{equation}\label{7Be-Delta}
\Delta_1(k^2(E))=(1+E/\epsilon)(a_0+a_1 E+ a_2 E^2),
\end{equation}
for both the $J^\pi=3/2^-$ and $J^\pi=1/2^-$ states of $^7\rm{Be}={^3\rm{He}}+{^4\rm{He}}$  with the corresponding binding energies $\epsilon_J$.

\begin{figure*}[thb]
\begin{center}
\parbox{12.0cm}{\includegraphics[width=12.0cm]{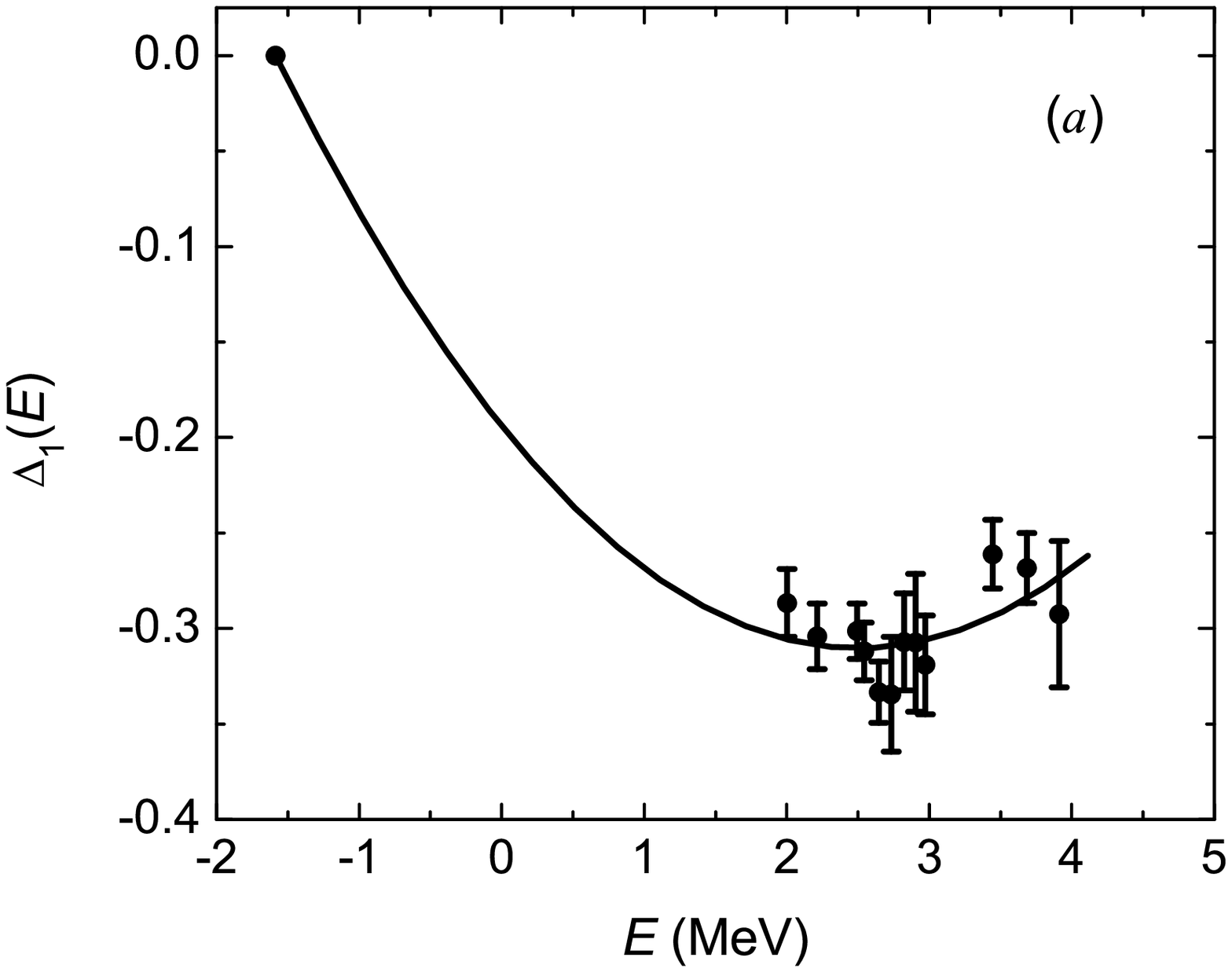}}
\parbox{12.0cm}{\includegraphics[width=12.0cm]{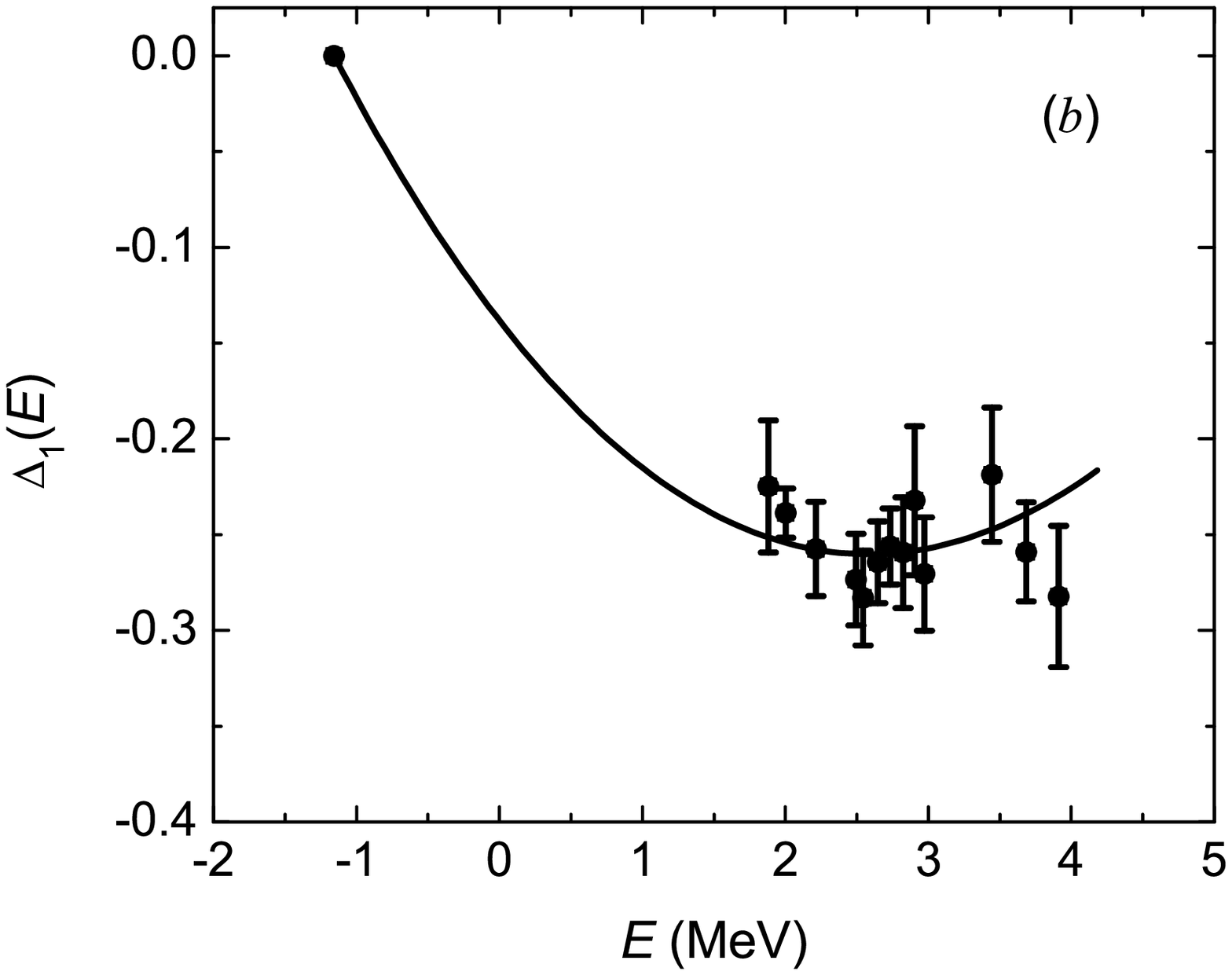}}
\caption{Dependence of $\Delta_1(k^2)$ (solid line) defined by Eq. (\ref{7Be-Delta}) on the c.m. energy for the $^3\rm{He}{^4\rm{He}}$ elastic collision in: (\textit{a}) $J^\pi =3/2^-$ state; (\textit{b}) $J^\pi =1/2^-$ state.  Experimental data (filled circles) obtained by using the experimental phase-shifts from Ref. \cite{boykin}.  \label{fig6}}
\end{center}
\end{figure*}

Fig. \ref{fig6} (\textit{a}) shows the dependence of the fitted $\Delta_1$-function on  the center-of-mass energy for the $^3\rm{He} ^4\rm{He}$ scattering  when $J^\pi =3/2^-$. For the fitting, we apply the experimental data from Ref. \cite{boykin} only. The calculated ANC is $3.389\pm 0.093$ if we apply the whole range of energies for the fitting. However, the results can vary greatly ($\sim 50\%$), if we change the energy range, or select the specific experimental data. It is clear that this is due to a lack of precision in the measurement of the phase-shifts. The same is true for the experimental data in the $J^\pi =1/2^-$ state (Fig. \ref{fig6} (\textit{b})).
The residues, NVCs, and ANCs are:
\begin{eqnarray}
J^\pi&=&3/2^-,\quad\epsilon=1.587\,\,\rm{MeV},\quad\rm{ground\,\, state:}\nonumber\\
\mid W_1\mid &=&7.137 \pm 0.398;\quad \mid G_1^2\mid=0.246 \pm 0.014,\rm{fm};\nonumber\\
&&\rm{ANC}=3.389 \pm 0.093\,\rm{fm}^{-1/2};\label{7Be-3-2}
\end{eqnarray}
\begin{eqnarray}
J^\pi&=&1/2^-,\quad\epsilon=1.157\,\,\rm{MeV},\quad\rm{excited\,\, state:}\nonumber\\
\mid W_1\mid &=&4.259 \pm 0.458;\quad \mid G_1^2\mid=0.125 \pm 0.013\,\rm{fm};\nonumber\\
&& \rm{ANC}=2.647 \pm 0.139\,\rm{fm}^{-1/2}.\label{7Be-1-2}
\end{eqnarray}
Fig. \ref{fig6} shows that the fitted curves can  deviate greatly from each other due to the larger uncertainties of the experimental data.
It is not surprising that the value of the ANC can vary widely if  we change the energy range. When we do this, the ANC in the state $J^\pi=3/2^-$ may be even less than in the state $J^\pi=1/2^-$.

\section{Conclusions}\label{conclusion}
In this paper we have shown that the new method is strictly valid for calculating the bound-state ANC in the frame of the effective-range expansion theory, which we call the $\Delta$-method. In fact, this constitutes a final step in the formulating the ERE approach, which simplifies calculations and excludes unnecessary details of the Coulomb interaction. In the standard ERE method, the total effective-range function must be fitted, and this function minus its Coulomb part has to be differentiated. The $\Delta$-method deals with the nuclear part, including the $\cot\delta$ multiplied by a factor (see Eq. (\ref{C02})) which depends on the Sommerfeld parameter and cancels out the essential zero energy singularity of the scattering phase shift. (see Eq. (6) in \cite{orlov16}).   Besides, we show  that $\Re h(\eta)$ given in the region $E>0$ continues  to $h(\eta)$ in the region $E<0$ and they both present the same  function in the real energy axis. This is an important part of our $\Delta$-method proof.

 As a result, we can use polynomial or Pad\'e approximations to continue the function $\Delta_l(k^2)$ analytically from the physical energy region to a bound-state pole where $\Delta_l$=0 together with Eq. (\ref{zero}).

This new $\Delta$-method is necessary when the product value $Z_1Z_2$ is large enough so that the nuclear term in the total effective-range function, which presents the sum of the nuclear and Coulomb terms, contributes only a little compared with the Coulomb term. Due to this, one can neither reproduce the experimental phase-shift data in the standard ERE method, nor obtain a proper value of the ANC. Furthermore, some bound poles may not be found in the standard calculations.

We have considered the $^{16}\rm{O}$ bound states, using as  input the elastic scattering phase-shift data from the R-matrix analysis in the channel $\alpha+{^{12}\rm{C}}$  for different $J^\pi$ states. This $\alpha{^{12}\rm{C}}$  system is a proper subject for the new $\Delta$-method application, because the nuclear term of the total effective-range function is very small compared with its Coulomb term.

Consequently, it is impossible to reproduce properly the $\alpha{^{12}\rm{C}}$  phase-shifts using the  ERE approach which we use in \cite{orlov16}. One result of the present work is the emergence of a second bound-state pole for the excited state $J^\pi=0^+$.  We did not find this state in \cite{orlov16} because the conventional method is not sensitive enough to nuclear properties. We  manage to calculate the properties of this state and other $^{16}\rm{O}$ bound states (see Table \ref{tab1}) including  the ground $J^\pi=0^+$ and sub-threshold bound states for $J^\pi=1^-$ and $2^+$. The Pad\'e approximation with the two poles is needed to describe the step behavior of the $\delta_2(E)$ phase shift. These poles are followed by two narrow resonances, owing to the $\delta_2(E)$ crossings $\pi/2$ and $3\pi/2$ values. These resonance positions are in good agreement with our calculation given in \cite{irgaz-14}.

We  also apply the $\Delta$-method to the ground $J^\pi=3/2^-$ and first excited $J^\pi=1/2^-$  states of $^7\rm{Be}$  treated as the bound states of the nuclei $^3\rm{He}$ and $^4\rm{He}$. Our results are compared with those published in the literature.

The possibility of using the condition $\Delta_l(k^2)$=0 for the ANC calculation is also considered in Ref. \cite{spren16}. But Ref. \cite{spren16} does not provide any proof of this condition, which we strictly derive in the present paper (see Section \ref{eff-range}). Since the $\Delta$-method leads to a polynomial function fitting or Pad\'e-approximant fitting, we introduce a bound-state energy into the input data by separating the corresponding bound-state factor in the formulas for the $\Delta_l(k^2)$ function (see Eqs. (\ref {Delta_0}) - (\ref{Delta_2})). We note that this is quite reasonable because the binding energy is known with  higher precision compared with the experimental phase-shift data.

The $\Delta$-method is applicable in any approach using the ERE to describe an elastic scattering amplitude. In the S-matrix method, designed for the description of resonances, (see \cite{irgaz-14}), the analytical continuation to the resonance pole is accomplished for the so-called 'potential' phase shift in the complex $k$ plane. We wish to point out that  the first attempt to use the ERE to calculate the ANC for resonances was achieved in  \cite{ENO}. In \cite{MIGOQ} the problem of calculating the resonance pole properties was solved using a similar S-matrix pole approach, but for a potential model.

The results of this paper are important for nuclear astrophysics studying new element creation in supernova explosions, as well as in the theory using Feynman diagrams to describe the amplitude of the direct nuclear reaction.

\section*{ACKNOWLEDGEMENTS}
This work was supported by the Russian Science Foundation  (Grant No. 16-02-0049).
We are grateful to H.\,M.~Jones for editing the English of this manuscript. B.F. Irgaziev thanks L.D. Blokhintsev for a useful discussion.

\end{document}